\documentclass[aps,showpacs,pre,superscriptaddress, twocolumn]{revtex4-1}
\newcommand{\be}{\begin{equation}}
\newcommand{\ee}{\end{equation}}
\newcommand{\bea}{\begin{eqnarray}}
\newcommand{\eea}{\end{eqnarray}}

\usepackage{epsfig}
\begin{document}
\title{Statistical Mechanics of a
Discrete Schr{\"o}dinger Equation with Saturable Nonlinearity}
\author{Mogens R. Samuelsen}
\address{Department of Physics, The Technical University of Denmark,
DK-2800 Kgs. Lyngby, Denmark}
\author{Avinash Khare}
\address{Indian Institute of Science Education and Research (IISER), Pune 411021, India}
\author{Avadh Saxena}
\address{Theoretical Division, Los Alamos National Laboratory,
Los Alamos, New Mexico, 87545, USA}
\author{Kim {\O}. Rasmussen}
\address{Theoretical Division, Los Alamos National Laboratory,
Los Alamos, New Mexico, 87545, USA}

\date{\today}


\begin{abstract}
We study the statistical mechanics of the one-dimensional discrete nonlinear Schr{\"o}dinger (DNLS) equation with saturable nonlinearity. Our study represents an
extension of earlier work [Phys. Rev. Lett. {\bf 84}, 3740 (2000)] regarding the statistical mechanics of the one-dimensional DNLS  equation with a cubic
nonlinearity. As in this earlier study we identify the spontaneous creation of localized excitations with a discontinuity in the partition function. The fact that this phenomenon 
is retained in the saturable DNLS is non-trivial, since in contrast to the cubic DNLS whose nonlinear character is enhanced as the excitation amplitude increases, the saturable DNLS in fact becomes increasingly linear as the excitation amplitude increases. We explore the nonlinear dynamics of this phenomenon by direct numerical simulations. 
\end{abstract}
\maketitle
During the past two decades much research has been devoted to nonlinear mechanisms
for the storage and transport of localized coherent packages
of energy and charge in spatially periodic systems, which may be described by discrete lattice models in one, two, or three spatial
dimensions \cite{rev, PGK_book, flach}. Generally such systems may support intrinsic localized modes, or discrete breathers, which are 
time-periodic, spatially localized solutions to the dynamical lattice equations \cite{mackay}. 

A much older interest \cite{FPU} in such nonlinear systems arises from their 
ability to thermalize through their intrinsic dynamics leading to equipartition of excitation energy among the linear modes. Localization and equipartition appears somewhat contradictory and  we have earlier 
explored their  interrelation  within the framework of the Discrete Nonlinear Schr{\"o}dinger (DNLS) equation with a cubic nonlinearity \cite{KOR1,KOR2}. We found that in the case of the cubic DNLS these phenomena occur in 
distinct parameter domains. Thermalization in the Gibbsian sense occurs in the low energy part of the parameter space while the nonlinear dynamics for higher energies lead to the spontaneous creation of 
localized excitations prohibiting thermalization. It has been argued \cite{rumpf} that the statistical process behind this localization leads towards an entropy maximization. The entropy maximization is driven by 
optimally distributing the effects of fluctuations (which carry the main part of the system entropy) among different conserved quantities. Localization is therefore absent (Gibbsian regime) if insufficient energy is supplied 
by the initial conditions. If a surplus of energy is provided by the initial conditions, it is allotted to the high amplitude structures (non-Gibbsian regime) that absorb large amounts of energy while using few lattice positions. 
The specific separation of these two regimes was derived analytically \cite{KOR1}. 

Extending this prior work we will here detail a similar study for  the DNLS with a saturable nonlinearity,
which physically describes, e.g., discrete spatial solitons in a tight-binding approximation of one-dimensional optical waveguide arrays made from photorefractive crystals \cite{lederer}. Another example where this equation arises is that of Bose-Einstein condensates in optical lattices \cite{BEC}.  Mathematically the saturable DNLS reduces to the 
cubic DNLS in the low amplitude (energy) limit, while it becomes essentially linear in the large amplitude limit. The saturable DNLS is known to support localized excitations \cite{krss1,krss2,melvin,si}, but it is unclear how the interplay of thermalization and spontaneous localization occurs in this case where strong localization essentially leads to linear dynamics. 

To explore these issues we
are concerned with the DNLS equation with a saturable nonlinearity
\begin{equation}
i\dot{\phi}_n + (\phi_{n+1}+\phi_{n-1})+
\frac{\nu|\phi_n|^2}{1+|\phi_n|^2}\phi_n=0,
\label{EQ:Field1}
\end{equation}
or equivalently through the simple gauge transformation $\phi_n=\psi_n \exp(i\nu t)$
\begin{equation}
i\dot{\psi}_n + (\psi_{n+1}+\psi_{n-1})-
\frac{\nu\psi_n}{1+|\psi_n|^2}=0,
\label{EQ:Field}
\end{equation}
where $\psi_n$ ($\phi_n$) is a
complex valued ``wave function" at site $n$ and $\nu$ is  a real parameter that controls the nonlinearity of the system. It is important to note that
for small excitation amplitudes $|\psi_n| \ll 1$ ($|\phi_n| \ll 1$) both equations reduce to the cubic DNLS.

Equation (\ref{EQ:Field}) represents a Hamiltonian system for the canonically conjugated pairs $i\psi_n$ and $\psi_n^*$ with the Hamiltonian
\bea
{\cal H} = \sum_{n=1}^{N}  \Big [(\psi_n^*\psi_{n+1}
+\psi_n\psi_{n+1}^*) - \nu \ln \left (1+|\psi_n|^2\right ) \Big],
\label{EQ:Hamiltonian}
\eea
so that Eq. (\ref{EQ:Field}) is given by
$i \dot \psi_n =-\frac{\partial{\cal H}}{\partial \psi_n^*}$.
In addition to the Hamiltonian, ${\cal H}$,
the quantity ${\cal A}=\sum_{n=1}^{N} | \psi_n|^2$ is also conserved by the dynamics of Eq. (\ref{EQ:Field}) and 
serves as the norm of the system.

Our objective is to calculate the grand-canonical partition function ${\cal Z}$ of this system and to accomplish this we first apply the  canonical transformation
$|\psi_n|^2=A_n$, $\psi_n=\sqrt{A_n}e^{i\theta_n}$,  which brings the Hamiltonian ${\cal H}$ of Eq. (\ref{EQ:Hamiltonian}) into the form
\bea
{\cal H}=\sum_{n=1}^N \Big[2\sqrt{A_n A_{n+1}}\cos(\theta_n-\theta_{n+1})-\nu\ln(1+A_n)\Big],
\label{EQ:Hamiltonian2}
\eea
and the norm becomes ${\cal A} =\sum_{n=1}^N A_n$.

In terms of these variables the  partition function ${\cal Z}$ is 

\be
Z=\prod_{n=1}^N\int_0^{2\pi} d\theta_n\int_0^\infty dA_n
e^{-\beta(\cal H+\mu\cal A)},
\label{EQ:Partitionfunction}
\ee
where the parameter $\mu$ is introduced analogous to a chemical potential to ensure the conservation of ${\cal A}$. From this partition function all thermodynamic quantities of the system can be calculated. We 
will in particular be interested in the averaged norm density $a=\langle {\mathcal A} \rangle /N$ and the averaged energy density $h=\langle {\mathcal H} \rangle /N$, which are given by the following expressions 
\be
a=-\frac{1}{\beta N} \frac{\partial \ln Z}{\partial \mu};  ~~~ h+\mu a= -\frac{1}{N}\frac{\partial \ln Z}{\partial \beta}.
\label{EQ:aandh}
\ee
In Eq. (\ref{EQ:Partitionfunction})
the integration over the phase variables $\theta_n$ can be preformed straightforwardly reducing the partition function to
\bea
&&\frac{Z}{(2\pi)^N}=\\ \nonumber &&\prod_{n=1}^N\int_0^\infty dA_nI_0 \left (2\beta\sqrt{A_nA_{n+1}} \right)
e^{-\beta\mu A_n}(1+ A_n)^{\beta\nu} , 
\label{EQ:Partitionfunction2}
\eea
where $I_0$ denotes the modified Bessel function.  Further treatment of the partition function has to be carried out numerically using for example the transfer integral approach \cite{KOR1}. However, 
in the two limits $\beta \rightarrow \infty$ and $\beta \rightarrow 0$ we can gain further analytical insight. Noticing that the Hamiltonian is bounded from below one 
can observe that its minimum (corresponding to $\beta \rightarrow \infty$) is realized by the plane wave $ \psi_n=\sqrt{ a} \exp in\pi$, whose energy density is $h=-2a-\nu\ln (1+a)$. 
When $\beta \rightarrow 0$ while $\beta \mu$ remains finite the modified Bessel function in the 
expression for ${\cal Z}$ can reasonably be approximated by unity, which leads to
\be
Z=(2\pi)^N\prod_{n=1}^N \int_0^\infty dA_n
e^{-\beta \mu A_n}(1+A_n)^{\beta\nu}.
\label{EQ:partitionfunction}
\ee
Making the substitution $1+A_n=x_n$ brings the partition function
$Z$ into
\be
Z=(2\pi)^Ne^{N\beta\mu}
E(\beta\nu,\beta\mu)^N,
\label{EQ:partitionfunction2}
\ee
where the function $E(x,y)$ is given by
\be
E(x,y)=\int_1^\infty du\,u^x\, e^{-yu} . 
\ee
Keeping $\beta\mu = \gamma$ finite as we take the limits $\beta \rightarrow 0$ and $\mu \rightarrow \infty$, we can, using
 Eq. (\ref{EQ:aandh}) and the properties for $E(x,y)$ listed in the appendix, obtain expressions for $a$ and $h$
\bea
a&=&-\frac{1}{\beta} \left (\beta +\beta\frac{\partial \ln E(0,y)}{\partial y} \Big |_{y=\gamma} \right )\\
&=&-1+\frac{E(1,\gamma)}{E(0,\gamma)}=
-1+1+\frac{1}{\gamma}=\frac{1}{\gamma}.
\eea
and
\bea
h+\mu a&=&-\mu-\nu\frac{\partial \ln E(x,\gamma)}{\partial x} \Big |_{x=0}-\mu \frac{\partial \ln E(0,y)}{\partial y} \Big |_{y=\gamma}\nonumber \\
&=&-\mu-\frac{\nu}{E(0,\gamma)}\frac{\partial E(x,\gamma)}{\partial x} \Big |_{x=0}+\mu \left (1+\frac{1}{\gamma} \right )\nonumber \\
&=&\frac{\mu}{\gamma}-\nu\frac{\gamma}{e^{-\gamma}}\frac{E_1(\gamma)}{\gamma}=\mu a -\nu e^{\gamma} E_1(\gamma)
\eea
so that
\be
h=-\nu\, e^\gamma \,E_1(\gamma)=-\nu\, e^{\frac{1}{a}}\,E_1\left (\frac{1}{a}\right ).
\label{line}
\ee
We can now compare with the result of the cubic DNLS by considering small $a$. Using the last equality of Eq. (\ref{expans}) in 
the limit of small $a$, Eq. (\ref{line}) becomes
\be
h \simeq-\nu \,e^{\frac{1}{a}}\, a \,e^{\frac{-1}{a}}\, (1-a)=-\nu a + \nu a^2.
\label{cubic}
\ee 
\begin{figure}[h]
\includegraphics[trim=1.7cm 6.4cm 2cm 6.5cm, clip=true,width=\columnwidth]{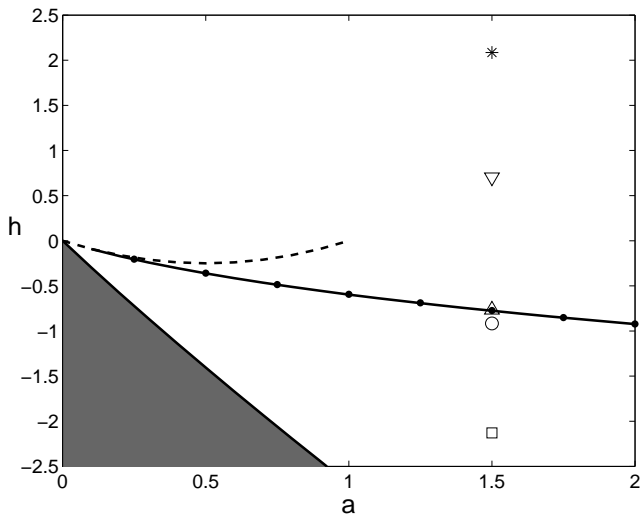}
\caption{\label{fig1} Parameter space ($a,h$) [for $\nu=1$], where the shaded area is inaccessible. The thick lines represent $\beta= 0$ [Eq.(\ref{line})] and $\beta=\infty$ [$h=-2a-\nu\ln (1+a)$]  and thereby bound the regime in which 
Gibbsian thermodynamics applies. The points are obtained numerically by applying the transfer integral technique to Eq. (\ref{EQ:Partitionfunction2}) using $\beta=10^{-8}$. The dashed line represents Eq. (\ref{cubic}) which corresponds to the cubic DNLS. Initial conditions for direct (micro-canonical) numerical calculations (see Figs. \ref{fig2}, \ref{fig3}, and \ref{fig4}) are marked by a square, a circle, an up-pointing triangle, a down-pointing triangle, and a star.}
\end{figure}
\begin{figure}[h]
\includegraphics[trim=1.3cm 6.4cm 2cm 6.5cm, clip=true,width=\columnwidth]{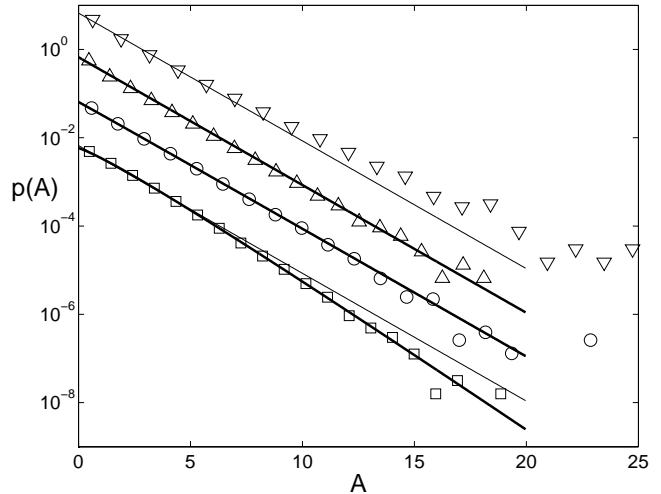}
\caption{\label{fig2} Distributions of $A= |\psi_n|^2$ for four different energies at $a=1.5$ [for $\nu=1$]. The symbols show data obtained by direct numerical solution.  Solid lines show the distribution function obtained by solution (using the transfer integral technique) of Eq. (\ref{EQ:Partitionfunction2}), while the thin lines show the function $\exp(-A/a)/a.$ The curves have been shifted vertically to facilitate visualization. All data is for $a=1.5$ while the energies are $h=-2.13$ (squares), $h=-0.97$ (circles), $h=-0.76$ (up-pointing triangles), and $h=0.70$ (down-pointing triangles).}
\end{figure}
In Fig. \ref{fig1} we illustrate, for $\nu=1$,  these results in the ($a,h$)-plane. The thick lines represent $\beta= 0$ [Eq. (\ref{line})] and $\beta=\infty$ [$h=-2a-\nu\ln (1+a)$]  that bound the regime in which 
Gibbsian thermodynamics applies. The region below the minimum energy line $\beta= \infty$, is of course inaccessible and is therefore shaded. However, the region above the $\beta=0$ line is accessible but cannot 
be captured by Gibbsian thermodynamics. In the corresponding region for the cubic DNLS we found \cite{KOR1} that the nonlinear dynamics lead to the spontaneous appearance of strongly 
localized discrete breather excitations. The dots
overlaid on the $\beta=0$ line represent values obtained numerically by applying the transfer integral technique (see details in Ref. \cite{KOR1}) to Eq. (\ref{EQ:Partitionfunction2}) using $\beta=10^{-8}$. The numerical 
solution does not ignore the modified Bessel function in Eq. (\ref{EQ:Partitionfunction2}) and it therefore directly verifies the validity of the expression in Eq. (\ref{line}). Finally, the dashed line illustrates the 
small amplitude approximation in Eq. (\ref{cubic}), which corresponds to the cubic DNLS \cite{cubicnote}. 
\begin{figure}
\includegraphics[trim=1.3cm 6.4cm 2cm 6.5cm, clip=true,width=\columnwidth]{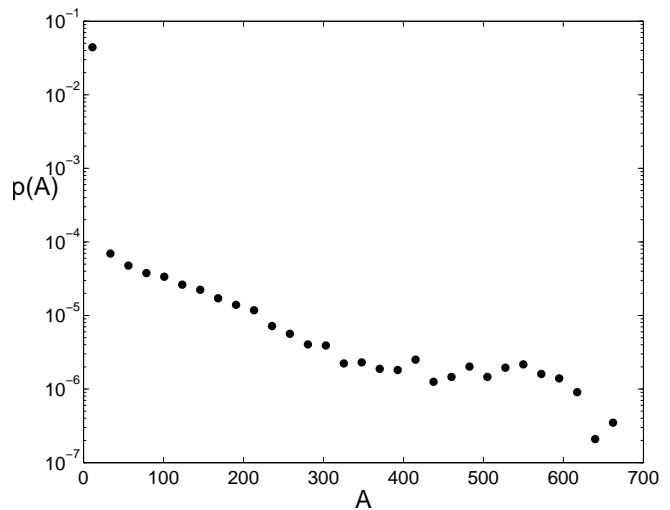}
\caption{\label{fig3}Distributions of $A= |\psi_n|^2$ for $h=2.09$ at $a=1.5$ [for $\nu=1$]. The distribution is obtained by direct numerical solution of Eq. (\ref{EQ:Field}).}
\end{figure}
In order to explore the differences in the dynamics in the two regions separated by the $\beta=0$ line we perform a number of direct numerical simulations of Eq.(\ref{EQ:Field}), with $\nu=1$. These numerical simulations represent a 
micro-canonical ensemble since they are performed to conserve both norm and energy. We use initial conditions of the form $\psi_n=\sqrt{a}\exp(i\theta n)$, which can be realized in both regimes by tuning the wave number $\theta$ and the norm $a$. The energy of these initial conditions is given by $h=2a\cos\theta-\ln(1+a)$. 

Our choices of initial conditions are marked in Fig. \ref{fig1} and the result of simulations with these initial conditions are shown in Figs. \ref{fig2} and \ref{fig3}. Figure {\ref{fig2} shows the amplitude distributions $p(A)$ resulting from long-time simulations. The numerically obtained distributions are shown with triangles, circles, and squares while the distribution functions obtained by the transfer integral approach are shown with thick lines. The thin lines show the function $\exp(-A/a)/a$. As the above analytical work demonstrates, this function approximates the distribution function in the limit of small
$\beta$. The validity of this approximation is clearly verified by Fig. \ref{fig2} since the thin lines are only discernible for $\beta$ sufficiently large. Figure \ref{fig2} also clearly demonstrates that the distribution arising from initial conditions above the $\beta =0$ line are significantly different from the Gibbsian distribution functions. While the Gibbsian distribution functions are concave the non-Gibbsian distribution functions are convex and have 
significant tails at large values of $A$. This convexity is reminiscent of the discussion in Ref. \cite{KOR1} regarding "negative temperatures" and is a subject that recently has been 
discussed in more detail \cite{new} in this context. These features are even more pronounced in the example given in Fig. \ref{fig3} which originates from initial conditions located  deep within the  non-Gibbsian regime. The long tail distributions  indicate the presence of a large amplitude excitations in the dynamics. The abrupt drop in the distribution function at very small excitation amplitudes clearly shows that the 
large amplitude excitations exist on an extensive background of small amplitude excitations.  This scenario is quite similar to Rumpf's \cite{rumpf} description of the dynamics of 
the cubic DNLS in this regime, which in turn suggests that also in this case does localized high-amplitude excitations absorb a surplus of energy when they emerge as a consequence of the production of entropy in the small ßuctuations.

In Fig. \ref{fig4} we directly illustrate these excitations with a spatio-temporal snapshot of the dynamics. A few very large amplitude excitations occur early in the dynamics.  These excitations are much larger in amplitude than what is typically observed in the cubic DNLS and although they are very discrete, only spanning 1-3 sites they are highly mobile \cite{si}. This mobility allows the excitations to collide and these collisions appear to lead to further amplitude enhancement. In spite of this high mobility the probability distribution function still reaches stationarity very slowly and require integration 
times above $10^6$ time units.The process remains slow because 
the collisions become quite rare as the excitation amplitudes increase. 
\begin{figure}[h]
\includegraphics[trim=1.1cm 6.4cm 1.3cm 8cm, clip=False,width=\columnwidth]{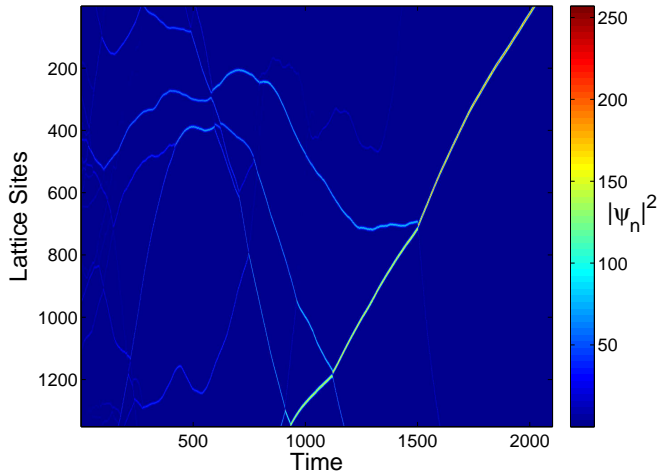}
\caption{(Color online) Spatio-temporal snapshot of the dynamics in the strongly non-Gibbssian regime for $h=2.09$ at $a=1.5$ [for $\nu=1$].\label{fig4} }
\end{figure}

In summary, we have studied a nonlinear Schr{\"o}dinger equation with saturable nonlinearity, and derived the separation between Gibbsian and non-Gibbsian regime in the ($a,h$) phase space. Through analytical calculations  supported by direct numerical simulations we have, as in the case of the cubic DNLS, been able to link the non-Gibbsian regime to the appearance of localized modes. The fact that this phenomenon 
is retained in the saturable DNLS is significant since in contrast to the cubic DNLS whose nonlinear character is enhanced as the excitation amplitude increases the saturable DNLS in fact becomes increasingly linear as the excitation amplitude increases.
\vspace{0cm}
\acknowledgments
This research was carried out under the auspices of the National Nuclear Security Administration of the US Department of Energy at Los Alamos National Laboratory under Contract No. DE-AC52-06NA25396.

\appendix*
\section{Properties of $E(x,y)$}
Some specific properties of the function $E(x,y)$ and its derivatives:
\bea
E(x,y)&=&\int_1^\infty u^xe^{-yu}du\\
&=&e^{-y}\int_0^\infty e^{-yz+x\ln(1+z)}~dz.
\nonumber
\eea
For the n'th derivative of $E(x,y)$ with respect to $x$ we find
\be
\frac{\partial^n E(x,y)}{\partial x^n}
=e^{-y}\int_0^\infty(\ln(1+z))^ne^{-yu+x\ln(1+z)}~dz,
\nonumber
\ee
and for the n'th derivative of $E(x,y)$ with respect to $y$ we find
\bea
\frac{\partial^n E(x,y)}{\partial y^n}&=&(-1)^n\int_1^\infty u^{x+n}e^{-yu+x\ln u}~du\\ \nonumber&=&(-1)^nE(x+n,y).
\eea
Using these expressions we find the following 
\be
E(0,y)=\frac{e^{-y}}{y}; ~~~  E(1,y)=\frac{e^{-y}}{y}\left (1+\frac{1}{y} \right )
\label{expans}
\ee
and 
\bea 
\frac{\partial E(x,y)}{\partial x}\Big |_{x=0}
&=&e^{-y}\int_0^\infty\ln(1+z)e^{-yu}~dz,\\
&=&e^{-y}\int_0^\infty \left (z-\frac{z^2}{2}\nonumber
+\frac{z^3}{3}-\cdot\cdot\cdot \right )e^{-yz}~dz.\\
&=&\frac{e^{-y}}{y^2} \left (1-\frac{1}{y}+\frac{2}{y^2}-\cdot\cdot\cdot \right  )= \frac{E_1(y)}{y}.\nonumber
\eea
The last expression follows for example from Eq. (5.1.51) of Ref. \cite{stegun} where the 
exponential integral is defined as $E_1(y)=\int^\infty_y \frac{e^{-t}}{t} dt$ (see Eq. 5.1.1 of Ref. \cite{stegun}).


\begin{thebibliography}{10}
\bibitem{cubicnote}When comparing these results to the results for the cubic DNLS presented in Ref. \cite{KOR1} it is important to note that 
Eq. (\ref{EQ:Field}) contains the additional  $-\nu\psi$ term in the small amplitude limit compared to the corresponding equation in Ref. \cite{KOR1}. This term accounts for the $-\nu a$ term in Eq. (\ref{cubic}).
\end{thebibliography}
\end{document}